\documentclass[preprintnumbers,amsmath,amssymbm,prd]{revtex4}
\usepackage{epsfig}
\usepackage{graphicx}
\usepackage{amssymb}

\begin{document}
\title{Charged massive scalar field configurations supported by a spherically symmetric charged reflecting shell}
\author{Shahar Hod}
\affiliation{The Ruppin Academic Center, Emeq Hefer 40250, Israel}
\affiliation{ }
\affiliation{The Hadassah Institute, Jerusalem 91010, Israel}
\date{\today}

\begin{abstract}
\ \ \ The physical properties of bound-state charged massive scalar
field configurations linearly coupled to a spherically symmetric
charged reflecting shell are studied {\it analytically}. To that
end, we solve the Klein-Gordon wave equation for a static scalar
field of proper mass $\mu$, charge coupling constant $q$, and
spherical harmonic index $l$ in the background of a charged shell of
radius $R$ and electric charge $Q$. It is proved that the
dimensionless inequality $\mu R<\sqrt{(qQ)^2-(l+1/2)^2}$ provides an
upper bound on the regime of existence of the composed
charged-spherical-shell-charged-massive-scalar-field configurations.
Interestingly, we explicitly show that the {\it discrete} spectrum
of shell radii $\{R_n(\mu,qQ,l)\}_{n=0}^{n=\infty}$ which can
support the static bound-state charged massive scalar field
configurations can be determined analytically. We confirm our
analytical results by numerical computations.
\end{abstract}
\bigskip
\maketitle

\section{Introduction}

The influential no-hair theorems \cite{NSH} have revealed the
interesting fact that spherically symmetric asymptotically flat
black holes cannot support static massive scalar field
configurations in their exterior regions \cite{Notensk,Hodrc,HerR}.
Motivated by this well known property of spherically symmetric
static black holes, we have recently \cite{Hodup} extended this
no-scalar-hair theorem to the regime of regular \cite{Notereg}
curved spacetimes. In particular, it was proved in \cite{Hodup} that
spherically symmetric compact reflecting \cite{Noteref} stars cannot
support regular self-gravitating neutral scalar field configurations
in their exterior regions.

One naturally wonders whether this no-scalar-hair behavior
\cite{Hodup} is a generic feature of compact reflecting objects? In
particular, we raise here the following physically intriguing
question: Can regular static {\it charged} massive scalar field
configurations be supported by a compact spherically symmetric
charged reflecting object? In order to address this interesting
question, in this paper we shall study, using {\it analytical}
techniques, the Klein-Gordon wave equation for a static linearized
scalar field of proper mass $\mu$ and charge coupling constant $q$
in the background of a spherically symmetric charged reflecting
shell of radius $R$ and electric charge $Q$.

Our results (to be proved below) reveal the fact that, for given
parameters $\{\mu,q,l\}$ \cite{Notell} of the charged massive scalar
field, there exists a {\it discrete} set of shell radii
$\{R_n(\mu,qQ,l)\}_{n=0}^{n=\infty}$ which can support the static
bound-state charged massive scalar field configurations. In
particular, as we shall explicitly show below, the regime of
existence of these composed
charged-spherical-shell-charged-massive-scalar-field configurations
is restricted by the characteristic inequality $(qQ)^2>(\mu
R)^2+(l+{1/2})^2$ \cite{Noteunit,Noteqm}. This relation implies, in
particular, that spatially regular bound-state configurations made
of neutral scalar fields \cite{Hodup} cannot be supported by a
spherically symmetric compact reflecting object.

\section{Description of the system}

We shall analyze the physical properties of a scalar field $\Psi$ of
proper mass $\mu$ and charge coupling constant $q$ which is linearly
coupled to a spherically symmetric charged shell of radius $R$. The
shell is assumed to have negligible self-gravity:
\begin{equation}\label{Eq1}
M,Q\ll R\  ,
\end{equation}
where $\{M,Q\}$ are the proper mass and electric charge of the
shell, respectively.

Decomposing the static scalar field $\Psi$ in the form \cite{Noteom}
\begin{equation}\label{Eq2}
\Psi(r,\theta,\phi)=\sum_{lm}e^{im\phi}S_{lm}(\theta)R_{lm}(r)\ ,
\end{equation}
one finds that the spatial behavior of the radial scalar
eigenfunctions $\{R_{lm}(r)\}$ in the spacetime region outside the
charged spherical shell is governed by the ordinary differential
equation \cite{HodPirpam,Stro,HodCQG2,Hodch1,Hodch2}
\begin{equation}\label{Eq3}
{{d} \over{dr}}\Big(r^2{{dR_{lm}}\over{dr}}\Big)+UR_{lm}=0\ ,
\end{equation}
where
\begin{equation}\label{Eq4}
U=(qQ)^2-(\mu r)^2-K_l\  .
\end{equation}
Here $K_l=l(l+1)$ is the characteristic eigenvalue of the angular
scalar eigenfunctions $\{S_{lm}(\theta)\}$ \cite{Heun,Abram}.

The bound-state (spatially localized) charged massive scalar field
configurations that we shall analyze below are characterized by
asymptotically decaying eigenfunctions
\begin{equation}\label{Eq5}
\Psi(r\to\infty)\sim {{1}\over{r}}e^{-\mu r}\
\end{equation}
at spatial infinity. In addition, the presence of the spherically
symmetric charged reflecting shell at $r=R$ dictates the boundary
condition
\begin{equation}\label{Eq6}
\Psi(r=R)=0\
\end{equation}
for the characteristic scalar eigenfunctions.

In the next section we shall explicitly show that the radial
differential equation (\ref{Eq3}), which determines the spatial
behavior of the characteristic radial eigenfunctions $\{R_{lm}(r)\}$
of the charged massive scalar fields in the background of the
charged spherical shell, is amenable to an {\it analytical}
treatment.

\section{The resonance equation for the composed charged-spherical-shell-charged-massive-scalar-field configurations}

As we shall now show, the characteristic radial equation (\ref{Eq3})
for the charged massive scalar eigenfunction $R_{lm}(r)$ can be
solved analytically. Defining the new radial function
\begin{equation}\label{Eq7}
\psi_{lm}=r^{1/2}R_{lm}\
\end{equation}
and using the dimensionless radial coordinate
\begin{equation}\label{Eq8}
z=\mu r\  ,
\end{equation}
one obtains the differential equation \cite{Noteomt}
\begin{equation}\label{Eq9}
z^2{{d^2\psi}\over{dz^2}}+z{{d\psi}\over{dz}}-\big[z^2+(l+{1\over2})^2-(qQ)^2\big]\psi=0\
\end{equation}
for the characteristic radial scalar eigenfunction $\psi$.

The general solution of the radial differential equation (\ref{Eq9})
can be expressed in terms of the modified Bessel functions (see Eq.
9.6.1 of \cite{Abram}) \cite{Notembf}:
\begin{equation}\label{Eq10}
\psi(z)=A\cdot K_{\nu}(z)+B\cdot I_{\nu}(z)\  ,
\end{equation}
where
\begin{equation}\label{Eq11}
\nu^2\equiv (l+{1\over2})^2-(qQ)^2\
\end{equation}
and $\{A,B\}$ are normalization constants. The asymptotic large-r
(large-z) behavior of the radial solution (\ref{Eq10}) is given by
(see Eqs. 9.7.1 and 9.7.2 of \cite{Abram})
\begin{equation}\label{Eq12}
\psi(z\to\infty)=A\cdot \sqrt{{{\pi}\over{2z}}}e^{-z}+B\cdot
{{1}\over{\sqrt{2\pi z}}}e^z\ .
\end{equation}
Taking cognizance of the boundary condition (\ref{Eq5}), which
characterizes the asymptotic spatial behavior of the bound-state
(spatially localized) scalar configurations, one deduces that the
coefficient of the exploding exponent in (\ref{Eq12}) must vanish:
\begin{equation}\label{Eq13}
B=0\  .
\end{equation}
One therefore concludes that the bound-state configurations of the
charged massive scalar fields in the background of the charged
spherical shell are characterized by the radial eigenfunction
\begin{equation}\label{Eq14}
\psi(r)=A\cdot K_{\nu}(\mu r)\  .
\end{equation}

Taking cognizance of Eq. (\ref{Eq14}) and the boundary condition
(\ref{Eq6}) which is dictated by the presence of the spherically
symmetric reflecting shell, one finds the characteristic resonance
equation
\begin{equation}\label{Eq15}
K_{\nu}(\mu R)=0\
\end{equation}
for the composed static
charged-spherical-shell-charged-massive-scalar-field configurations.
Interestingly, as we shall show below, the resonance condition
(\ref{Eq15}) determines the {\it discrete} set of shell radii
$\{R_n(\mu,qQ,l)\}_{n=0}^{n=\infty}$ which can support the
bound-state charged massive scalar field configurations.

In the next section we shall prove that the resonance condition
(\ref{Eq15}) can only be satisfied in the bounded regime
\begin{equation}\label{Eq16}
(qQ)^2>(\mu R)^2+(l+{1\over2})^2\  .
\end{equation}
The necessary inequality (\ref{Eq16}), to be proved below, implies
in particular that spatially regular static bound-state
configurations made of neutral scalar fields cannot be supported by
a spherically symmetric compact reflecting object.

\section{The domain of existence of the charged massive scalar hair}

Using the boundary conditions (\ref{Eq5}) and (\ref{Eq6}), one concludes that the scalar eigenfunction $\psi$,
which characterizes the radial behavior of the charged massive
scalar fields, must have (at least) one extremum point,
$z=z_{\text{peak}}$, outside the spherically symmetric charged
reflecting shell. In particular, at this extremum point the radial
scalar eigenfunction $\psi$ is characterized by the relations
\begin{equation}\label{Eq17}
\{{{d\psi}\over{dz}}=0\ \ \ \text{and}\ \ \
\psi\cdot{{d^2\psi}\over{dz^2}}<0\}\ \ \ \ \text{for}\ \ \ \
z=z_{\text{peak}}\  .
\end{equation}

Substituting (\ref{Eq17}) into (\ref{Eq9}), one finds the
characteristic inequality
\begin{equation}\label{Eq18}
z^2_{\text{peak}}+(l+{1\over2})^2-(qQ)^2<0\  .
\end{equation}
Taking cognizance of (\ref{Eq8}) and using the inequality
$r_{\text{peak}}>R$, one finds from (\ref{Eq18}) that the composed
charged-spherical-shell-charged-massive-scalar-field configurations
are characterized by the inequality (\ref{Eq16}). In particular,
this inequality sets the upper bound
\begin{equation}\label{Eq19}
\mu R<\sqrt{(qQ)^2-(l+{1\over2})^2}\
\end{equation}
on the radius of the central charged supporting shell.

For later purposes, it is important to point out that the inequality
(\ref{Eq19}) [or equivalently, the inequality (\ref{Eq16})] implies
that the static charged massive scalar field configurations are
characterized by the relation $\nu^2<0$ [see Eq. (\ref{Eq11})],
which implies
\begin{equation}\label{Eq20}
i\nu\in\mathbb{R}\  .
\end{equation}

\section{Bound-state resonances of the charged massive scalar fields in the background of the charged spherical shell}

The analytically derived resonance condition (\ref{Eq15}) for the
composed static charged-spherical-shell-charged-massive-scalar-field
configurations can easily be solved numerically. In particular, one
finds that, for given parameters $\{\mu,q,l\}$ of the charged
massive scalar field, there exists a {\it discrete} set of shell
radii,
\begin{equation}\label{Eq21}
\cdots R_2<R_1<R_0\equiv R^{\text{max}}< {{|\nu|}\over{\mu}}\  ,
\end{equation}
which can support the static bound-state charged massive scalar
field configurations. In Table \ref{Table1} we display the {\it
largest} possible dimensionless radius $\mu R^{\text{max}}(\nu)$ of
the supporting charged shell for various values of the dimensionless
physical parameter $\nu$ \cite{NoteWol}. From the data presented in Table
\ref{Table1} one finds that $\mu R^{\text{max}}(\nu)$ is a
monotonically increasing function of $|\nu|$. Note, in particular,
that the values of the dimensionless supporting radii $\mu
R^{\text{max}}(\nu)$ conform to the analytically derived upper bound
(\ref{Eq19}).

\begin{table}[htbp]
\centering
\begin{tabular}{|c|c|c|c|c|c|c|c|}
\hline $\nu\ $ & \ $\ 1i\ $\ \ & \ $\ 20i\ $\ \ & \
$\ 40i\ $\ \ \ & \ $\ 60i\ $\ \ & \ $\ 80i\ $\ \ & \ $\ 100i\ \ $\ \ \\
\hline
\ $\ \ \mu R^{\text{max}}(\nu)\ $\ \ \ &\ 0.0640\ \ &\ 15.343\ \ &\ 33.955\ \ &\ 52.999\ \ &\ 72.244\ \ &\ 91.609\ \ \\
\hline
\end{tabular}
\caption{Composed
charged-spherical-shell-charged-massive-scalar-field configurations.
We display, for various values of the dimensionless physical
parameter $\nu$ [see Eq. (\ref{Eq11})], the largest possible
dimensionless radius $\mu R^{\text{max}}(\nu)$ of the charged
reflecting shell which can support the static bound-state charged
massive scalar field configurations. One finds that $\mu
R^{\text{max}}(\nu)$ is a monotonically increasing function of
$|\nu|$. Note, in particular, that the dimensionless radii of the
charged supporting shells are bounded from above by the inequality
$\mu R^{\text{max}}(\nu)<|\nu|$ [see Eqs. (\ref{Eq11}) and
(\ref{Eq19})].} \label{Table1}
\end{table}

As we shall now show, the characteristic resonance equation
(\ref{Eq15}) for the composed static charged-shell-charged-field configurations
can be solved {\it analytically} in the asymptotic regimes
\cite{Notecomp1}
\begin{equation}\label{Eq22}
\mu R\ll1\
\end{equation}
and \cite{Notecomp2,Notemunu,Notesch,Schw1,Schw2}
\begin{equation}\label{Eq23}
\mu R\gg1\  .
\end{equation}
To this end, we first note that the resonance condition (\ref{Eq15})
can be expressed in the from (see Eq. 9.6.2 of \cite{Abram})
\begin{equation}\label{Eq24}
I_{\nu}(\mu R)=I_{-\nu}(\mu R)\  .
\end{equation}

\subsection{The regime $\mu R\ll1$ of small charged supporting shells}

Using the small argument ($\mu R\ll1$) approximation
\begin{equation}\label{Eq25}
I_{\nu}(z)\simeq {{(z/2)^{\nu}}\over{\Gamma(\nu+1)}}\ \ \ \
\text{for}\ \ \ \ z\to 0\
\end{equation}
of the modified Bessel function (see Eq. 9.6.7 of \cite{Abram}), one
finds from (\ref{Eq24}) the characteristic resonance equation
\cite{Notegam}
\begin{equation}\label{Eq26}
({1\over2}\mu R)^{2\nu}=-{{\Gamma(\nu)}\over{\Gamma(-\nu)}}\
\end{equation}
for the dimensionless physical quantity $\mu R$ in the small-radius
regime (\ref{Eq22}). From Eqs. (\ref{Eq11}) and (\ref{Eq26}) one
finds the expression \cite{Notem1,Notegamr}
\begin{equation}\label{Eq27}
\mu
R(n)=2\Big[{{\Gamma(\nu)}\over{\Gamma(-\nu)}}\Big]^{1/2i|\nu|}\times
e^{-\pi(n+{1\over2})/|\nu|}\ \ \ ; \ \ \ n\in\mathbb{Z}
\end{equation}
for the {\it discrete} spectrum of shell radii which can support the
static bound-state charged massive scalar field configurations in
the small-radius regime (\ref{Eq22}).

\subsection{The regime $\mu R\gg1$ of large charged supporting shells}

Using the uniform asymptotic expansion
\begin{equation}\label{Eq28}
I_{\nu}(\nu z)\simeq
{{e^{\nu\eta}}\over{\sqrt{2\pi\nu}(1+z^2)^{1/4}}}\ \ \ \text{with}\
\ \ \eta=\sqrt{1+z^2}+\ln[z/(1+\sqrt{1+z^2})]\ \ \ \ \text{for}\ \ \
\ |\nu|\to\infty\
\end{equation}
of the modified Bessel function (see Eqs. 9.7.7 and 9.7.11 of
\cite{Abram}), one finds from (\ref{Eq24}) the characteristic
resonance equation
\begin{equation}\label{Eq29}
{{xe^{\sqrt{1-x^2}}}\over{1+\sqrt{1-x^2}}}=(-i)^{1/2i|\nu|}
\end{equation}
for the physical quantity $\mu R$ in the large-radius regime
(\ref{Eq23}), where here we have used the dimensionless parameter
\begin{equation}\label{Eq30}
x\equiv {{\mu R}\over{|\nu|}}\  .
\end{equation}

Substituting \cite{Noteeps}
\begin{equation}\label{Eq31}
x=1-\epsilon\ \ \ \ \text{with}\ \ \ \ \ 0<\epsilon\ll1\
\end{equation}
into (\ref{Eq29}), one obtains the leading order equation
\cite{Notemunu,Notemi}
\begin{equation}\label{Eq32}
1-{{4\sqrt{2}}\over{3}}\cdot\epsilon^{3/2}+O(\epsilon^{5/2})=1-{{\pi(1+4n)}\over{2|\nu|}}+O(|\nu|^{-2})\
\ \ ; \ \ \ n=0,1,2,...\
\end{equation}
for $\epsilon$, which yields
\begin{equation}\label{Eq33}
\epsilon=\Big({{3\pi}\over{8\sqrt{2}}}\cdot{{1+4n}\over{|\nu|}}\Big)^{2/3}\
.
\end{equation}
Taking cognizance of Eqs. (\ref{Eq30}), (\ref{Eq31}), and
(\ref{Eq33}), one finds
\begin{equation}\label{Eq34}
\mu
R(n)=|\nu|\Big[1-\Big({{3\pi}\over{8\sqrt{2}}}\cdot{{1+4n}\over{|\nu|}}\Big)^{2/3}\Big]\
\ \ ; \ \ \ n=0,1,2,...
\end{equation}
for the {\it discrete} spectrum of shell radii which can support the
static bound-state charged massive scalar field configurations in
the large-radius regime (\ref{Eq23}).

\section{Numerical confirmation}

It is of physical interest to verify the validity of the
analytically derived formulas (\ref{Eq27}) and (\ref{Eq34}) for the
{\it discrete} set of charged shell radii which can support the
static bound-state charged massive scalar field configurations. In
Table \ref{Table2} we display the dimensionless discrete radii $\mu
R^{\text{analytical}}$ of the charged reflecting shells as obtained
from the analytically derived formula (\ref{Eq27}) in the
small-radius regime (\ref{Eq22}). We also display the corresponding
dimensionless radii $\mu R^{\text{numerical}}$ of the charged shells
as obtained from a direct numerical solution of the characteristic
resonance equation (\ref{Eq15})  \cite{NoteWol}. One finds a remarkably good
agreement \cite{Noteapg} between the approximated radii of the
charged supporting shells [as calculated from the analytically
derived formula (\ref{Eq27})] and the corresponding exact radii of
the charged shells [as obtained numerically from the analytically
derived resonance equation (\ref{Eq15})].

\begin{table}[htbp]
\centering
\begin{tabular}{|c|c|c|c|c|c|c|c|}
\hline \text{Formula} & \ $\mu R(n=-1)$\ \ & \ $\mu R(n=0)$\ \ & \
$\mu R(n=1)$\ \ & \ $\mu R(n=2)$\ \ & \ $\mu R(n=3)$\ \ & \ $\mu R(n=4)$\ \ \\
\hline
\ {\text{Analytical}}\ [Eq. (\ref{Eq27})]\ \ &\ 2.2888\ \ &\ 1.2210\ \ &\ 0.6514\ \ &\ 0.3475\ \ &\ 0.1854\ \ &\ 0.0989\\
\ {\text{Numerical}}\ [Eq. (\ref{Eq15})]\ \ &\ 2.4245\ \ &\ 1.2393\ \ &\ 0.6541\ \ &\ 0.3479\ \ &\ 0.1855\ \ &\ 0.0989\\
\hline
\end{tabular}
\caption{Composed
charged-spherical-shell-charged-massive-scalar-field configurations
with $\nu=5i$ [see Eq. (\ref{Eq11})]. We present the dimensionless
discrete radii $\mu R$ of the central charged reflecting shells as
calculated from the analytically derived formula (\ref{Eq27}). We
also present the corresponding dimensionless radii of the central
charged shells as obtained from a direct numerical solution of the
characteristic resonance equation (\ref{Eq15}). One finds a
remarkably good agreement \cite{Noteapg} between the approximated
radii of the charged reflecting shells [as calculated from the
analytically derived formula (\ref{Eq27})] and the corresponding
exact radii of the charged shells [as obtained numerically from the
analytically derived resonance equation (\ref{Eq15})].}
\label{Table2}
\end{table}

In Table \ref{Table3} we present the dimensionless discrete radii
$\mu R^{\text{analytical}}$ of the charged reflecting shells as
deduced from the analytically derived formula (\ref{Eq34}) in the
large-radius regime (\ref{Eq23}). We also present the corresponding
dimensionless radii $\mu R^{\text{numerical}}$ of the charged shells
as obtained from a direct numerical solution of the resonance
equation (\ref{Eq15}) \cite{NoteWol}. Again, one finds a remarkably good agreement
between the approximated radii of the charged reflecting shells [as
calculated from the analytically derived formula (\ref{Eq34})] and
the corresponding exact radii of the charged shells [as obtained
numerically from the analytically derived resonance condition
(\ref{Eq15})].

\begin{table}[htbp]
\centering
\begin{tabular}{|c|c|c|c|c|c|c|c|}
\hline \text{Formula} & \ $\mu R(n=0)$\ \ & \ $\mu R(n=1)$\ \ & \
$\mu R(n=2)$\ \ & \ $\mu R(n=3)$\ \ & \ $\mu R(n=4)$\ \ & \ $\mu R(n=5)$\ \ \\
\hline
\ {\text{Analytical}}\ [Eq. (\ref{Eq34})]\ \ &\ 194.82\ \ &\ 184.86\ \ &\ 177.60\ \ &\ 171.37\ \ &\ 165.77\ \ &\ 160.59\\
\ {\text{Numerical}}\ [Eq. (\ref{Eq15})]\ \ &\ 189.32\ \ &\ 181.56\ \ &\ 175.36\ \ &\ 169.99\ \ &\ 165.17\ \ &\ 160.75\\
\hline
\end{tabular}
\caption{Composed
charged-spherical-shell-charged-massive-scalar-field configurations
with $\nu=200i$ [see Eq. (\ref{Eq11})]. We display the dimensionless
discrete radii $\mu R$ of the central charged reflecting shells as
calculated from the analytically derived formula (\ref{Eq34}). We
also display the corresponding dimensionless radii of the charged
supporting shells as obtained from a direct numerical solution of
the resonance condition (\ref{Eq15}). One finds a remarkably good
agreement between the approximated radii of the charged supporting
shells [as calculated from the analytically derived formula
(\ref{Eq34})] and the corresponding exact radii of the charged
shells [as obtained numerically from the analytically derived
resonance condition (\ref{Eq15})].} \label{Table3}
\end{table}

\section{Summary and Discussion}

The elegant no-hair theorems \cite{NSH} have revealed the remarkable
fact that asymptotically flat black holes cannot support static
scalar field configurations in their exterior regions.
Interestingly, this result has recently been extended \cite{Hodup}
to the regime of spherically symmetric regular \cite{Notereg} curved
spacetimes. In particular, it was proved in \cite{Hodup} that
neutral self-gravitating static scalar field configurations cannot
be supported in the exterior spacetime region of a spherically
symmetric compact reflecting \cite{Noteref} star.

Following the intriguing no-scalar-hair property observed in
\cite{Hodup} for compact reflecting objects, we have raised here the
following physically interesting question: Can a spherically
symmetric {\it charged} reflecting shell support regular static {\it
charged} massive scalar field configurations in its exterior region?
In order to address this intriguing question, in the present paper
we have solved {\it analytically} the Klein-Gordon wave equation for
a linearized static scalar field of mass $\mu$ and charge coupling
constant $q$ in the background of a spherically symmetric charged
reflecting shell of radius $R$ and electric charge $Q$.

The main results derived in this paper and their physical
implications are as follows:

(1) We have proved that the upper bound [see Eqs. (\ref{Eq11}),
(\ref{Eq19}), and (\ref{Eq20})]
\begin{equation}\label{Eq35}
\mu R<|\nu|\ \ \ \ ; \ \ \ \ i\nu\equiv
\sqrt{(qQ)^2-(l+{1\over2})^2}\in\mathbb{R}\
\end{equation}
on the dimensionless radius of the charged reflecting shell provides
a necessary condition for the existence of composed static
charged-spherical-shell-charged-massive-scalar-field configurations.

(2) It was shown that, for given physical parameters $\{\mu,q,l\}$
of the charged massive scalar field, there exists a {\it discrete}
set of charged shell radii $\{R_n(\mu,qQ,l)\}_{n=0}^{n=\infty}$
which can support the static spatially regular bound-state charged
massive scalar field configurations. In particular, we have proved
that the characteristic discrete set of supporting shell radii is
determined by the analytically derived compact resonance condition
$K_{\nu}(\mu R)=0$ [see Eq. (\ref{Eq15})].

(3) We have explicitly shown that the physical properties of the
composed static charged-shell-charged-massive-scalar-field
configurations can be studied {\it analytically} in the asymptotic
regimes of small [see Eq. (\ref{Eq22})] and large [see Eq.
(\ref{Eq23})] charged supporting shells. In particular, we have
provided compact analytical formulas for the discrete spectra of
supporting shell radii in the asymptotic regimes $\mu R\ll1$
\cite{Notecomp1} and $\mu R\gg1$ \cite{Notecomp2} [see Eqs.
(\ref{Eq27}) and (\ref{Eq34}), respectively].

(4) Former analytical studies have revealed the interesting fact
that spherically symmetric {\it neutral} reflecting stars share the
no-scalar-hair property with asymptotically flat black holes
\cite{NSH,Hodup}. On the other hand, the results derived in the
present paper have revealed the fact that compact {\it charged}
reflecting objects have a much richer phenomenology. In particular,
while the elegant no-hair theorems of Bekenstein and Mayo
\cite{BekMay} (see also \cite{Hodtp}) have shown that spherically
symmetric asymptotically flat charged black holes cannot support
static charged massive scalar field configurations, in the present
work we have explicitly proved that charged reflecting shells {\it
can} support static bound-state charged massive scalar field
configurations in their exterior regions.

\bigskip
\noindent
{\bf ACKNOWLEDGMENTS}
\bigskip

This research is supported by the Carmel Science Foundation. I would
like to thank Yael Oren, Arbel M. Ongo, Ayelet B. Lata, and Alona B.
Tea for helpful discussions.



\begin{thebibliography}{99}
\bibitem{Notensk} It is worth mentioning that it has recently been
proved \cite{Hodrc,HerR} that non-spherically symmetric rotating
black holes can support stationary (rather than static) regular
massive scalar field configurations in their exterior regions.
\bibitem{Notereg} Here the term `regular spacetimes' refers to curved
spacetimes which possess no event horizons.
\bibitem{Noteref} A reflecting star is defined as a compact regular object for which
the external scalar field vanishes on its surface [see Eq.
(\ref{Eq6}) below].
\bibitem{Notell} Here $l$ is the spherical harmonic index of the charged massive
scalar field [see Eq. (\ref{Eq2}) below].
\bibitem{Noteunit} We shall use natural units in which $G=c=\hbar=k_{\text{B}}=1$.
\bibitem{Noteqm} Note that the physical parameters $\{\mu.q\}$ of the charged massive scalar field
stand for $\{\mu/\hbar,q/\hbar\}$. Hence, these physical parameters
have the dimensions of $($length$)^{-1}$.
\bibitem{Noteom} The integer indices $m$ and $l$ (with $l\geq |m|$) are respectively the
azimuthal harmonic parameter and the spherical harmonic parameter of
the charged massive scalar field mode.
\bibitem{Noteomt} We shall henceforth omit the angular harmonic indices $\{l,m\}$ for brevity.
\bibitem{Notembf} Here $I_{\nu}(z)$ and $K_{\nu}(z)$ are respectively the
modified Bessel function of the first kind and the modified Bessel
function of the second kind \cite{Abram}.
\bibitem{NoteWol} The numerically computed roots of the modified Bessel function $K_{\nu}(x)$ [see the analytically derived resonance equation (\ref{Eq15})]
were obtained using the WolframAlpha: Computational Knowledge Engine.
\bibitem{Notemunu} Note that the strong inequality $\mu R\gg1$ corresponds to
$|\nu|\gg1$ [see Eqs. (\ref{Eq11}) and (\ref{Eq19})].
\bibitem{Notesch} It is worth noting that the Schwinger quantum pair-production mechanism
\cite{Schw1} in the electric field of the charged shell is
exponentially suppressed in the $E_{\text{shell}}\ll
E_{\text{c}}\equiv \mu^2/q\hbar$ regime \cite{Schw1,Schw2} (here
$E_{\text{shell}}=Q/R^2$ is the electric field at the surface of the
charged shell). Taking cognizance of the inequality (\ref{Eq19}),
one deduces that this quantum effect is exponentially suppressed in
the $\mu R<|qQ|\ll (\mu R)^2$ regime. These two inequalities imply
the strong inequality $\mu R\gg1$.
\bibitem{Notem1} Here we have used the relation $-1=e^{-i\pi(1+2n)}$,
where the resonance parameter $n$ is an integer.
\bibitem{Notegamr} Taking cognizance of the relation (\ref{Eq20}) and using Eq. 6.1.23
of \cite{Abram}, one deduces that the ratio
${{\Gamma(\nu)}/{\Gamma(-\nu)}}$ is a complex number of the form
$e^{i\phi}$, where $\phi\in\mathbb{R}$. This implies
$[{{\Gamma(\nu)}/{\Gamma(-\nu)}}]^{1/2i|\nu|}\in\mathbb{R}$. One
therefore concludes that, for imaginary values of the dimensionless
parameter $\nu$ [see Eqs. (\ref{Eq11}) and (\ref{Eq20})], the
dimensionless radii $\mu R(n)$ of the charged supporting shells as
obtained from the analytically derived formula (\ref{Eq27}) are
purely real numbers.
\bibitem{Noteeps} Note that $\mu R<|\nu|$ [see Eqs. (\ref{Eq11}) and (\ref{Eq19})],
which implies $x<1$ (or equivalently, $\epsilon>0$).
\bibitem{Notemi} Here we have used the relation $-i=e^{-i\pi({1\over2}+2n)}$,
where the resonance parameter $n$ is an integer.
\bibitem{Noteapg} Interestingly, from the data presented in Table
\ref{Table2} one finds that the agreement between the numerical data
[as obtained from a direct numerical solution of the resonance
equation (\ref{Eq15})] and the analytical formula (\ref{Eq27}) is
quite good even in the regime $\mu R=O(1)$. This finding is quite
surprising since the analytically derived formula (\ref{Eq27}) for
the discrete spectrum of charged shell radii which can support the
static bound-state charged massive scalar field configurations is
formally valid in the restricted regime $\mu R\ll1$ [see Eqs.
(\ref{Eq22}) and (\ref{Eq25})].
\end{thebibliography}
\end{document}